\documentclass[letterpaper, 10 pt, conference]{ieeeconf}  

\IEEEoverridecommandlockouts                              

\usepackage{amsmath} 
\usepackage{amssymb}
\usepackage{graphicx}

\title{\LARGE \bf Defense via Behavior Attestation against Attacks in Connected and Automated Vehicles based Federated Learning Systems}

\usepackage{algpseudocode}
\usepackage{algorithm}

\begin{document}

\title{Ordered-logit pedestrian stress model for traffic flow with automated vehicles}


\author{ \parbox{3 in}{\centering Kimia Kamal\\
         Laboratory of Innovations in Transportation\\
         Ryerson University\\
         Toronto, Canada\\
         {\tt\small kimia.kamal@ryerson.ca}}\vspace*{0.25cm}
         \hspace*{ 0.5 in}
         \parbox{3 in}{ \centering Bilal Farooq\\
        Laboratory of Innovations in Transportation\\
        Ryerson University\\
         Toronto, Canada\\
         {\tt\small bilal.farooq@ryerson.ca}}\\
         \parbox{3 in}{ \centering Mahwish Mudassar\\
                 Laboratory of Innovations in Transportation\\
        Ryerson University\\
         Toronto, Canada\\\vspace*{0.25cm}
         {\tt\small mahwish.mudassar@ryerson.ca}}
        \hspace*{ 0.5 in}
        \parbox{3 in}{ \centering Arash Kalatian\\
        Institute of Transport Studies\\
        University of Leeds\\
         Leeds, UK\\
         {\tt\small a.kalatian@leeds.ac.uk}}\\
         Cite as: \underline{Kamal, K., Farooq, B., Mudassar, M., Kalatian, A. (2022) Ordered-logit pedestrian  stress model}\\
         \underline{for traffic flow with automated vehicles. In: IEEE Intelligent Vehicles Symposium}\\
         \underline{Workshops (XXIV Workshops), 2022, Aachen, Germany.}
}

\maketitle

\begin{abstract}
An ordered-logit model is developed to study the effects of Automated Vehicles (AVs) in the traffic mix on the average stress level of a pedestrian when crossing an urban street at mid-block. Information collected from a galvanic skin resistance sensor and virtual reality experiments are transformed into a dataset with interpretable average stress levels (low, medium, and high) and geometric, traffic, and environmental conditions. Modelling results indicate a decrease in average stress level with the increase in the percentage of AVs in the traffic mix. 
\end{abstract}

\section{Introduction}
In the near future, fully automated vehicles (AVs) are expected to operate on urban roads. In dense urban areas, they must interact with other road users, including the pedestrians. For their sustainable adoption, safe operations, and features development, it is important that the pedestrian behaviour when interacting with AVs is well understood before their mass deployment. An important aspect of this effort is to investigate how the stress level of the pedestrian changes when interacting with an AV in different geometric, traffic, and environmental conditions.

For this purpose, we used the data from a range of scenarios developed in a dynamic and immersive Virtual Reality (VR) setup, where the participants crossed at mid-block on an urban street in Toronto, Canada. The pedestrian stress was measured using a Galvanic Skin Resistance (GSR) sensor. The raw temporal data were processed into interpretable and comparable average stress levels (low, medium, and high). Using the resulting dataset, an ordinal-logit model was developed to systematically quantify and analyze socio-demographics and the effects of different geometric, traffic mix, and environmental conditions on pedestrian stress level.

The rest of the paper is organized as follows. Existing literature on stress modelling and pedestrian-AV interaction is discussed in Section \ref{back}. The developed model is presented in Section \ref{meth}, while Section \ref{dat} presents the data and processing method. Detailed discussion on the estimation results is presented in Section \ref{res} and conclusions in Section \ref{con}.

\section{Background}
\label{back}
Quy and Kubiak \cite{quy1974comparison} in their seminal work in 1973, used GSR sensor data to conclude that as a participant goes through the same task several times, their average stress level decreased. Furthermore, it took much higher effort to raise their stress level for the same task. This observation can be extended to transportation, where it is expected that individuals whose primary mode is walking would be less stressed in crossing the street. Zeedyk and Kelly \cite{zeedyk2003behavioural} analyzed the behaviour of children-adult pairs crossing the street. It was reported that young males held adult's hands less often when crossing the road compared to females. This observation provided an anecdotal evidence that males were less stressed when crossing than female. 
Several simulator studies concluded that younger people were more likely to take risks when crossing the road and felt less stressed \cite{meir2015child, charron2012child}. Kitabayashi et al. \cite{kitabayashi2015analysis} developed a naturalistic experiment in which pedestrians were fitted with GSR and heart rate sensors and were asked to walk along a route. The study concluded that pedestrians were more stressed when crossing roads with higher level of traffic.

Due to the absence of a considerable number of fully Automated Vehicles (AVs) on roads, their interaction with pedestrian is currently studied in laboratory settings, using tools like VR \cite{kalatian2021decoding}. Open access sensors data from Waymo, Lyft, and other AV companies have been made available, but they lack the revealed or measured behaviour of the pedestrians. In terms of pedestrian stress analysis, Mudassar et al. \cite{mudassar2021analysis} used GSR sensor with experiments in VR to measure the change in pedestrian stress when crossing the road in various conditions. The study concluded that the instantaneous relative stress of the pedestrian increased as the distance of an approaching vehicle decreased. The study also found positive correlation between instantaneous stress and acceleration of the pedestrian.

In this study we are interested in the effect of AVs in the flow, geometric, sociodemographic, and environmental conditions on the average stress level of the pedestrian when crossing the road. The resulting model can be useful to the policy makers and manufacturers in the sustainable adoption of AVs on urban roads. 

\section{Methodology}
\label{meth}
An ordinal regression model is widely used to predict a category from a ranked discrete choice set in which the labels of categories follow a relative ordering scale. In 1980, McCullagh first proposed multivariate extensions of generalized linear models for modeling ordinal, among which Proportional Odds Models (POM) or Ordered Logit Models is the most popular ordinal classification method \cite{mccullagh1980regression}. This model is considered between regression and multi-class classification such that an unobserved continuous variable, known as a latent variable, relates to ordinal responses through thresholds. The basic structure of ordered logit model or threshold models for a set of ordinal responses, $r_1<r_2<\cdots< r_k$ can be mathematically represented as follows: 

\begin{equation}
U^*_n=\beta x_n + \varepsilon_n 
\end{equation}

\begin{align*} 
U_n = r_k \quad \textrm{if} \quad \delta_{k-1}<U^*_n<\delta_k \qquad \textrm{for}\quad k=1,2,...K
\end{align*}

Where $U_n$ represents the ordinal response and $U^*_n$ is the latent variable, which is assumed to be a linear function of explanatory variables $(x_n)$ with associated coefficients $\beta$, which remain constant between ordinal categories. In addition, $\varepsilon_n$ is the random error term of utility function following logistic distribution. It is of note that in the ordinal regression model, the effect of variables on ordinal categories is assumed to be constant across response categories, leading to the parallel regression assumption. In addition to regression coefficients, a set of thresholds $(\delta_k)$ associating the latent variable to an ordinal response is estimated based on the data. These thresholds must satisfy the constraint $\delta_1\leq\delta_2\leq\cdots\leq\delta_k$. 
This model is based on the cumulative probabilities of the response variable; therefore, in contrast to other discrete choice models like Multinominal Logit Model (MNL) used for classification problems, the ordered logit model are not directly expressed in terms of probabilities of the ordinal categories. In other words, the choice probability of choosing $k^{th}$ category over an ordinal choice set is equals to the difference of cumulative probabilities as follows: 

\begin{equation}
\begin{split}
P(U_n=r_k) & = P(U^*_n>\delta_k)-P(U^*_n>\delta_{k-1}) \\
 & = \frac{1}{\textrm{exp}(\beta x_n-\delta_k)}-\frac{1}{\textrm{exp}(\beta x_n-\delta_{k-1})}
\end{split}
\end{equation}

Since inherently there is a natural ordinal relationship among different responses of individuals' stress in a dangerous situation, in this study, we use an ordinal logit model for evaluating pedestrians stress levels in different traffic situations involving automated vehicles. In other words, in the real world, individual stress levels are usually categorized into natural ranked categories like low, medium, and high. Therefore, the ordered logit model can be an appropriate structure for modelling pedestrians stress levels.
In this study, the Maximum Likelihood method is employed to estimate the set of optimal parameters and develop a predictive ordered logit model from the data.

\section{Data}
\label{dat}
In this study, we use a dataset from Mudassar et al. \cite{mudassar2021analysis} on pedestrian's stress level when interacting with normal and automated vehicles collected by using Virtual Immersive Reality Environment (VIRE) technology. VIRE is a virtual reality simulation framework providing a perception of being physically present in futuristic scenarios by immersing the participants in an artificial 3D environment \cite{farooq2018virtual}. Therefore, VIRE technology provides an opportunity for researchers to evaluate pedestrians crossing behaviour in the face of unprecedented traffic conditions. Mudassar et al. analyzed the temporal changes in an individual pedestrian's stress levels in various scenarios defined based on different  controlled factors such as rules and regulations, street characteristics, automated vehicle features, traffic demand and environmental conditions \cite{mudassar2021analysis}. Galvanic Skin Response (GSR) sensors technology was used to measure the relative stress levels of a participant. The GSR sensor uses a small electrical charge to measure the amount of sweat an individual has on their finger \cite{mudassar2021analysis}. The greater the charge, the greater the sweat. Detailed information about the definition of scenarios can be found in their study.
As the GSR sensor only measures the relative change in the stress of a pedestrian compared to their initial stress, the data used in Mudassar et al. cannot be used directly to compare the stress among the pedestrians or develop prediction models. 
Therefore, in this study, we first computed the mean, minimum, and maximum stress levels for each pedestrian in a crossing scenario. The mean stress level of the pedestrian in a scenario was normalized according to their minimum and maximum stress level observed each scenario. Then, the Jenks Natural Breaks classification method is applied to categorize the normalized stress into low, medium, and high levels \cite{jiang2013head}. The discretization ensures a more interpretable and quantifiable analysis of the effects of various factors. The chosen method \cite{jiang2013head} classifies a given data into several groups in such a way as to minimize the variance of members of each group while maximizing the variance between different groups. The three discrete groups are obtained: low: pedestrians who relatively feel minor stress level, less than 0.4, in comparison to others, medium: pedestrians whose normalized mean stress level are between 0.4 to to 0.6, and high: pedestrians who are under stress of higher than 0.6.

\section{Results}
\label{res}
In Table \ref{table2}, the results obtained from the estimated ordered logit model are presented. As the main aim of this study is to analyze effective factors on pedestrians stress level in the presence of AVs, in the following, the impact of each explanatory variable is explained in detail.
Regarding the coefficient of explanatory variables, a positive coefficient shows that the related variable positively affects the probability of higher stress level and for a negative coefficient, the reverse is true. 
Furthermore, based on the values of thresholds parameters and the utility function, we are able to predict the category of pedestrians stress. For instance, if the utility function or latent variable is higher than $\delta_1$ and less than $\delta_2$, the pedestrians stress level is predicted to be medium. Some parameters were kept in the model despite their low t-statistics values because of their importance and the consistency of estimates.
It is of note that in this study, 180 participants experienced several different street scenarios during their participation. Thus the resulting 732 observations can be grouped by 180 participants, forming a panel data. In the estimated model, we did not account for the panel effect, but this can be investigated in future work.

\begin{table}[h]
\caption{Parameter estimates for ordered-logit model}
\label{table2}
\begin{center}
\renewcommand{\arraystretch}{1.3}
\begin{tabular}{ccc}
\hline
Variables & Coefficient & t-stats \\ \hline
\textbf{\emph{Traffic condition}} \\
Mixed traffic condition &	-0.621 & -1.172 \\
Fully automated condition & -0.325 & -0.823 \\\hline
\textbf{\emph{Road attributes}} \\
Two way with a median & 0.373 & 2.124 \\
Two way & 0.406 & 2.270 \\\hline
\textbf{\emph{Socio-demographic}} \\
Female & 1.000 & 5.081 \\
Age 40-49 & 2.002 & 2.452 \\
Age over 50 & 2.037 & 6.848 \\
Driving license & 1.138 & 3.721 \\
One car & 0.428 & 1.997 \\
Over one car &  0.413 & 1.832 \\
Public mode & 1.115 & 5.684 \\
Active mode & 1.003 & 4.148	\\
Walk to work or shop & -0.418 & -1.848\\\hline
\textbf{\emph{Environmental condition}} \\
Snowy & 0.238 & 1.514	 \\\hline
$\delta_1$ &\multicolumn{2}{c}{1.399 (2.296)}\\\hline
$\delta_2$& \multicolumn{2}{c}{3.268 (11.624)}\\\hline
No. observation	 & \multicolumn{2}{c}{732}\\\hline
No. parameters	 & \multicolumn{2}{c}{16} \\\hline
Initial Log-likelihood & \multicolumn{2}{c}{1032.57}\\\hline
Final Log-likelihood & \multicolumn{2}{c}{728.53}\\\hline
\end{tabular}
\end{center}
\end{table}

\subsection{Traffic condition}
Based on the information of Table \ref{table2}, the mixed traffic conditions and fully automated situation both decrease pedestrians stress levels compared to the current traffic condition in which pedestrians only face human-driven vehicles. This result highlights the importance of general trust in AVs. As in the experiments, participants could distinguish human-driven from AVs \cite{kalatian2021decoding}, and people generally have trust in braking systems of AVs, mixed and fully automated traffic condition likely make pedestrians more relaxed. In other words, this result demonstrates that the more trust in AVs leads to the less stress level and also underscores the effect of educational measures before the introduction of AVs into the roads. It is worth mentioning that as the main aim of our study is to show the effect of mixed traffic condition on pedestrians stress levels, the non-significant fully automated condition variable is not eliminated from the model due to its logical impact. In fact, our result may be rooted in our data, however, we consider the impact of this variable in the utility function to show the negative effect of this variable on pedestrians stress level for further research. 

\subsection{Road attributes}
The significant positive coefficient of road types shows that crossing from a two-way road or even a two-way road with a medium makes pedestrians more stressed compared to a one-way road. In other words, with the introduction of AVs to roads, pedestrians likely find the one-way roads safer for crossing and thus their stress level is getting lower. This result indicates the vital importance of urban design, affecting pedestrians behaviour, in the presence of AVs. In this study, the impact of density and lane width were insignificant, so they are not part of the final model.

\subsection{Socio-demographic}
Regarding individuals' characteristics, females may be under more stress when they decide to cross a road in the presence of AVs. Broadly, there is a stress gap between men and women and higher levels of stress is usually reported for women. Regarding pedestrians behaviour and their stress level, our study shows that a similar result may be observed in future traffic condition. For evaluating the effect of age group on pedestrians stress level, the 18-29 group is settled as a baseline category. The result of our model obtained from the VR data highlight that both 40-49 age group and pedestrians aged over 50 find the road condition more stressful in comparison to the baseline category. It is worth mentioning that the reverse impact was obtained for 30-39 age group, despite statistically insignificant effect. Furthermore, having a driving license causes higher stress levels. In fact, the more pedestrians are aware of driving rules and probable dangerous situations, the more they feel stressed. In addition, individuals who have access to private cars in their household considerably have more stress when they are waiting on the sidewalk and crossing the street. This result indicates that the greater chance of using a private car leads to greater stress level, rooted in their less experience in crossing the street. 
Our developed model significantly estimates the effect of individuals' lifestyles on their stress level. Based on the results, the probability of having higher stress considerably decreases if pedestrians usually walk to work or shop. In fact, if people daily take walking at least once, walking habit become a daily habit of this group and they become more familiar with crossing, resulting in lower stress. A reverse result is obtained for the impact of using public transportation. As public transportation users usually use different public modes for their daily activities, transition between different modes and fixed schedule of public transportation generally engenders more stress and tension in pedestrians. Our result indicates that in the face of AVs, pedestrians who prefer to regularly use public transportation will probably find the future condition more stressful. In a previous work, we found that in the presence of AVs, public transportation users highly likely wait longer than people using a private car \cite{kamalFarooq_2022}. Interestingly, this longer waiting time may be rooted in their higher stress level. Similarly, using active modes, including walking and bicycle increases the probability of feeling a higher level of stress. This positive impact may trace back to the strong impact of biking. In fact, in the real world, people who have a preference towards biking, usually feel more stress compared to pedestrians and other road users, due to less appropriate facilities and conditions for biking in urban roads. 

\subsection{Environmental condition}
Another noteworthy this virtual reality data is an opportunity allowing us to evaluate the effect of weather condition on pedestrians stress level in the future road conditions, although simulation of the real effect of lighting and weather condition has been still considered as a limitation of virtual reality experiments. According to our result, Snowy weather increases the probability of a higher stress level, since harsh weather condition afflict pedestrians' vision. In Toronto's context snow is an important variables, but depending on the location, other environmental conditions, e.g., rain, fog, and high winds, can also be included in future studies.

\section{Conclusion}
\label{con}
An ordered-logit model is developed to quantify and analyze the effects of AVs in the traffic flow on the average stress level of a pedestrian when crossing at mid-block on an urban street. The estimated parameters revealed that the average stress level of a pedestrian decreased progressively as the percentage of AVs in the traffic mix increased. This phenomena is perhaps due to the higher level of trust a pedestrian associates with the automatic breaking system, compared to a human driver's capacity to react and break. Adverse weather conditions, i.e. snowy weather, increased the average stress level of a pedestrian, so as the presence of two-way traffic. Having a median slightly decreased the stress. Participants whose main mode was walking experienced a lower stress level. Female and older participants experienced a higher stress level. We expect that these findings will be useful to both policy makers and AV manufacturers in developing more sustainable and acceptable policies, features, and solutions.

As a next step, the group behaviour of pedestrians when interacting with AVs can be studied. Another future direction could be to study the effects of different treatments that can substitute the absence of driver-pedestrian eye-contact in AVs on pedestrian behaviour. 

\bibliographystyle{IEEEtran}

\bibliography{Reference}

\begin{thebibliography}{10}
\providecommand{\url}[1]{#1}
\csname url@samestyle\endcsname
\providecommand{\newblock}{\relax}
\providecommand{\bibinfo}[2]{#2}
\providecommand{\BIBentrySTDinterwordspacing}{\spaceskip=0pt\relax}
\providecommand{\BIBentryALTinterwordstretchfactor}{4}
\providecommand{\BIBentryALTinterwordspacing}{\spaceskip=\fontdimen2\font plus
\BIBentryALTinterwordstretchfactor\fontdimen3\font minus
  \fontdimen4\font\relax}
\providecommand{\BIBforeignlanguage}[2]{{%
\expandafter\ifx\csname l@#1\endcsname\relax
\typeout{** WARNING: IEEEtran.bst: No hyphenation pattern has been}%
\typeout{** loaded for the language `#1'. Using the pattern for}%
\typeout{** the default language instead.}%
\else
\language=\csname l@#1\endcsname
\fi
#2}}
\providecommand{\BIBdecl}{\relax}
\BIBdecl

\bibitem{quy1974comparison}
R.~J. Quy and E.~W. Kubiak, ``A comparison between “aware” and “naive”
  conditions in the suppression of gsr activity,'' \emph{The Quarterly Journal
  of Experimental Psychology}, vol.~26, no.~4, pp. 561--565, 1974.

\bibitem{zeedyk2003behavioural}
M.~S. Zeedyk and L.~Kelly, ``Behavioural observations of adult--child pairs at
  pedestrian crossings,'' \emph{Accident Analysis \& Prevention}, vol.~35,
  no.~5, pp. 771--776, 2003.

\bibitem{meir2015child}
A.~Meir, T.~Oron-Gilad, and Y.~Parmet, ``Are child-pedestrians able to identify
  hazardous traffic situations? measuring their abilities in a virtual reality
  environment,'' \emph{Safety science}, vol.~80, pp. 33--40, 2015.

\bibitem{charron2012child}
C.~Charron, A.~Festoc, and N.~Gu{\'e}guen, ``Do child pedestrians deliberately
  take risks when they are in a hurry? an experimental study on a simulator,''
  \emph{Transportation research part F: traffic psychology and behaviour},
  vol.~15, no.~6, pp. 635--643, 2012.

\bibitem{kitabayashi2015analysis}
H.~Kitabayashi, X.~Zhang, Y.~Asano, and M.~Yoshikawa, ``An analysis of the
  walking environmental factors affecting the stress of pedestrians for route
  recommendation,'' in \emph{2015 16th IEEE International Conference on Mobile
  Data Management}, vol.~2.\hskip 1em plus 0.5em minus 0.4em\relax IEEE, 2015,
  pp. 44--49.

\bibitem{kalatian2021decoding}
A.~Kalatian and B.~Farooq, ``Decoding pedestrian and automated vehicle
  interactions using immersive virtual reality and interpretable deep
  learning,'' \emph{Transportation research part C: emerging technologies},
  vol. 124, p. 102962, 2021.

\bibitem{mudassar2021analysis}
M.~Mudassar, A.~Kalatian, and B.~Farooq, ``Analysis of pedestrian stress level
  using gsr sensor in virtual immersive reality,'' \emph{Collective Dynamics},
  vol.~6, 2022.

\bibitem{mccullagh1980regression}
P.~McCullagh, ``Regression models for ordinal data,'' \emph{Journal of the
  Royal Statistical Society: Series B (Methodological)}, vol.~42, no.~2, pp.
  109--127, 1980.

\bibitem{farooq2018virtual}
B.~Farooq, E.~Cherchi, and A.~Sobhani, ``Virtual immersive reality for stated
  preference travel behavior experiments: A case study of autonomous vehicles
  on urban roads,'' \emph{Transportation research record}, vol. 2672, no.~50,
  pp. 35--45, 2018.

\bibitem{jiang2013head}
B.~Jiang, ``Head/tail breaks: A new classification scheme for data with a
  heavy-tailed distribution,'' \emph{The Professional Geographer}, vol.~65,
  no.~3, pp. 482--494, 2013.

\bibitem{kamalFarooq_2022}
K.~Kamal and B.~Farooq, ``Interpretable ordinal-reslogit for pedestrian wait
  time modelling in virtual reality,'' in \emph{Transportation Research Board},
  2022, pp. 1--18.

\end{thebibliography}

\end{document}